\documentclass[11pt]{article}
\usepackage{graphicx,cite,epsfig}
\newcommand{\ba}{\begin{array}}
\newcommand{\ea}{\end{array}}
\newcommand{\bd}{\begin{displaymath}}
\newcommand{\ed}{\end{displaymath}}
\newcommand{\be}{\begin{equation}}
\newcommand{\ee}{\end{equation}}
\def\bt{\begin{table}}
\def\et{\end{table}}
\def\bi{\begin{itemize}}
\def\ei{\end{itemize}}
\def\bea{\begin{eqnarray}}
\def\eea{\end{eqnarray}}

\def\tq{\tilde{q}}

\def\g{\gamma}

\def\G{\Gamma}

\def\s{\sigma}

\def\slash {\!\!\!\!/}

\catcode`@=11 
\def \gsim{\mathrel{\mathpalette\@versim>}}
\def \lsim{\mathrel{\mathpalette\@versim<}}
\def \@versim#1#2{\lower0.4ex\vbox{\baselineskip\z@skip\lineskip\z@skip
     \lineskiplimit\z@\ialign{$\m@th#1\hfil##\hfil$%
     \crcr#2\crcr\sim\crcr}}}
\catcode`@=12 
\begin{document}
\begin{flushright}
HRI-P-05-09-001 \\
\end{flushright}
\begin{center}
{\Large\sc   Identifying the contributions of Universal Extra
Dimensions in the Higgs sector at linear $e^+ e^-$ colliders} \\
\vspace*{0.2in}
{\large\sl Anindya Datta$^{1,2~\dagger}$ {\rm and} Santosh Kumar Rai$^{1~\ddagger~\ast}$} \\
\vspace*{0.2in}
$^1$Harish-Chandra Research Institute, Chhatnag Road, Jhunsi, Allahabad, India 211 019.\\ 
$^2$Department of Physics, University of Calcutta, 92 A.P.C. Road, Kolkata, India 700 009.\\
\vspace*{0.5in}
{\Large\bf ABSTRACT}
\end{center}
\noindent 
We study the dilepton-dijet signal in the dominant Higgs production 
channel at a linear $e^+ e^-$ collider. We estimate the effects of 
Universal Extra Dimension (UED) by a simple analysis of the 
cross-sections. The heavy Kaluza-Klein excitations of the Standard Model
fields in UED can significantly alter the decay properties of the Higgs 
boson to loop-driven final states. We show that by taking a simple ratio 
between cross-sections of two different final states this difference
can be very easily highlighted. 
\noindent
\rm\normalsize

\vfill
\noindent$^{\dagger}$adphys@caluniv.ac.in ~~~~
\noindent$^{\ddagger}$santosh.rai@helsinki.fi \\
\noindent$^{\ast}$ {\sl \underline {Present address}:} \\
 High Energy Physics Division, Department of Physical Sciences, University
 of Helsinki,\\ and Helsinki Institute of Physics, P.O. Box 64,
 FIN-00014 University of Helsinki, Finland\\  
\newpage

\section{Introduction}
Theories of large extra spacetime dimensions \cite{LED1,LED2} and warped
extra dimensions \cite{RS1,RS2} have received a great deal of attention
during the past several years.  The most popular amongst them have
been the models which consider localization of Standard Model (SM)
fields on a 3-brane. In these scenario the SM fields can only feel the
effects of the extra dimensions through interactions with the
Kaluza-Klein (KK) excitations of the gravitational field in the bulk,
projected onto the brane by the usual KK mechanism.  However, within
some other models motivated from the framework of string theory
\cite{antoniadis1}, other non-gravitational fields can also propagate
in the bulk provided they do not disturb the experimental constraints,
which puts a limit on the maximum size of the extra dimension.

\bigskip\noindent
One such scenario, referred to as the Universal Extra Dimension (UED)
model \cite{appel}, allows all the Standard Model (SM) fields to propagate 
in extra dimensions. At tree level, the momentum along the extra
dimensions is also conserved, requiring pair production of the
Kaluza-Klein (KK) modes at colliders and preventing tree level mixing
effects from altering precision electroweak measurements. 
Values of compactification scale, as low as 250-300 GeV are allowed for 
one extra dimension \cite{appel-yee}.  
The phenomenological implications of UED have been extensively studied 
in the literature \cite{appel,phenolow1,phenolow2,phenolow3,phenolow4,
phenolow5,phenolow6,phenolow7,phenohigh1,phenohigh2,phenohigh3,phenohigh4,
phenohigh5,phenohigh6,phenohigh7,phenohigh8,phenohigh9,phenohigh10,phenohigh11,
phenohigh12,phenohigh13}.

\bigskip\noindent
Direct detection of UED KK states at future
colliders requires them to be pair produced due to the KK number
conservation and hence already puts a limit on the minimum energy at
which the collider should run to produce these particles. However,
effects of UED excitations can be also be observed in loops at much
lower energies \cite{phenolow1,phenolow2,phenolow3,phenolow4,phenolow5,
phenolow6,phenolow7} than that required to produce them on-shell. 
It is therefore interesting to determine whether there are other, indirect
ways in which the effects of UED can be detected at future high energy
colliders. One such possibility is through the modification of Higgs
decay properties due to the KK states. 

\bigskip\noindent
Higgs boson discovery will prove to be a crucial ingredient towards
understanding the mechanism of electroweak symmetry breaking.
Once discovered, a major goal would be
to determine its other intrinsic properties, couplings and its total
width with high accuracy in a model independent way.
The partial width of the Higgs decaying to the massless gauge boson is
of special interest, since there are no tree level couplings of the
Higgs to them and any contribution is generated at the one-loop level.
The di-photon partial width gets contribution through massive charged
particles in the loops while the gluon-gluon partial width gets
contributions from the heavy quarks running in the loops. The effective
loop induced couplings of $H\g\g$ and $Hgg$ are sensitive to new
contributions from particles which appear in various extensions of the
SM. Not only do these decay modes provide for a possible probe of new
physics particles which are too heavy to be produced directly but they
are also sensitive to scales far beyond the Higgs mass.
The partial decay widths for $H\to gg$,
$H\to\g\g$ and $H\to\g Z$ decay modes which are driven by loops can be
substantially modified due to KK excited modes of SM particles running
in the loops.  As pointed out in Ref.\cite{petriello} there is a
remarkably significant enhancement in the partial decay width of the
Higgs in $H\to gg$ due to the excited top quark loops. This can
greatly enhance the Higgs production at the Large Hadron Collider
(LHC) viz. the $gg \to H$ mode of production. However, there are 
large uncertainties associated with the PDF's and the choice of scale
(scale uncertainties)\cite{pdfscale1,pdfscale2} at the LHC which might just 
mask this enhancement and make it difficult to differentiate the contributions
coming from UED. Also the very fact that the $H\to gg$ or 
decays into light quarks cannot be observed at the LHC due to the huge 
QCD backgrounds, makes it difficult to look for UED contributions in the 
dijet final state. 
It is quite evident that one requires a linear electron-positron collider to 
study the properties of the Higgs boson, viz. its coupling to SM particles and 
its decay properties. In view of whether any new physics beyond the SM 
affects its properties, will be a crucial detail one needs to understand. 
The future lepton colliders would be 
instrumental in identifying the Higgs properties more precisely due to the much
cleaner environment. In this work we try to point out a very robust and simple
technique of identifying the effect of the contributions of the heavy KK-modes 
to the loop induced decay modes of the Higgs boson. In fact the technique is 
useful in testing the sensitivity of the scale of new physics beyond SM. 
For our study we focus on the UED model and we look at the
dominant mode of Higgs production at a future $e^+e^-$ collider and
its subsequent decay into two jets and try to identify the contribution
coming from UED. In Section~2 we give a very brief overview about the
model in consideration and how the partial width for $H\to gg$ gets
modified. In Section~3 we present our results and conclusions.

\section{The Model} 
The UED model, in its simplest form \cite{appel}, has all the SM particles
propagating in a single extra dimension, which is
compactified on an $S_1/Z_2$ orbifold with $R$ as the radius of
compactification.  Conservation of KK number
which is a consequence of momentum conservation along the extra
dimension forces the KK particles to be pair produced. Consequently, UED 
predicts a stable lightest Kaluza-Klein particle (LKP) which would be much
like the lightest supersymmetric particle (LSP). The LKP is a viable
candidate for dark matter \cite{LKP1,LKP2}.
Bulk and brane radiative effects \cite{branebulk1,branebulk2,
branebulk3,branebulk4} however break KK number down to a
discrete conserved quantity, the so called KK parity, $(-1)^n$, where
$n$ is the KK level. KK parity conservation in turn, implies that the
contributions to various precisely measured low-energy observables
only arise at loop level and are small \cite{phenolow1,phenolow2,phenolow3,
phenolow4,phenolow5, phenolow6,phenolow7}. As a result, the limits on the
scale of the extra dimension, from precision electro-weak data, are
rather weak, constraining $R^{-1}$ to be larger than 300 GeV.

\bigskip\noindent
The KK tower resulting on the four dimensional space-time has a tree
level mass given by
\bea
 m_n^2 = m^2 + \frac{n^2}{R^2}
\eea
where $n$ denotes the $n^{th}$-level of the KK tower and $m$
corresponds to the mass of the SM particle in question. This implies a
mass degeneracy in the $n^{th}$-level of the spectrum at least for the 
leptons and lighter quarks. This degeneracy is however
removed due to radiative corrections to the masses \cite{branebulk1,branebulk2,
branebulk3,branebulk4}. In the following, we use 
the tree-level masses, but our result is still valid with the
mass degeneracies lifted.

\bigskip\noindent
In this work, we are mainly interested in the modification of the
partial decay width of the Higgs in $H\to gg$ due to the KK tower of the
top quark running in the loops. Since the KK number is not violated at
any of the vertices inside a loop, the contributions come from all the
KK-excitations, with a decoupling nature for the higher modes. So 
only the first few KK modes have relevant contribution. Below we give
the contributions to the two-gluon decay width of the Higgs boson in the
SM and UED.

\bigskip\noindent
Since the $H \to gg$ proceeds through diagrams containing fermion
triangle loops and the coupling is proportional to the zero-mode mass of the
fermion even in the case of UED, we consider contributions of the KK
tower of the top quark only. The partial decay width for $H \to gg$ with 
UED contribution is \cite{petriello},
\bea 
\G(H\to gg) = \frac{G_F~m_H^3}{36\sqrt{2}\pi}
\left(\frac{\alpha_s(m_H)}{\pi}\right)^2~|I_q+\sum_n {\tilde I}_{\tq^{(n)}}|^2
\eea 
where $G_F$ is the Fermi constant, $\alpha_s(m_H)$ is the running 
QCD coupling evaluated at $m_H$  and $I_q, {\tilde I}_{\tq^{(n)}}$ are the
contributions of the loop integrals for the SM and UED case
respectively. These loop integral functions can be written down as:
\bea
I_q &=& \frac{3}{4}\lambda_q \left[2-(1-\lambda_q) F(\lambda_q)\right] 
 \nonumber \\
\tilde{I}_{\tq^{(n)}} &=& \frac{6}{4} \lambda_q \left[2-(1-\lambda_{\tq^{(n)}}) 
                      F(\lambda_{\tq^{(n)}}) \right]
\label{eqn:fnloop}
\eea
where $\lambda_i=4 m_i^2/M_H^2$. The $q$ stands for the different quark 
flavours in the SM while $\tq^{(n)}$ represents the $n^{th}$ KK excitation of 
the particular quark flavour. The function $F(\lambda)$ in the above 
expression is given by 
\bea
 F (\lambda)
& = & -2 \left(\sin^{-1} \frac{1}{\sqrt{\lambda}}\right)^2 ~~{\rm for}~~ \lambda \geq 1 \nonumber \\
& = & -\frac{\pi^2}{2} + \frac{1}{2}\log^2
\frac{1 + \sqrt{1 - \lambda}} {1 - \sqrt{1 - \lambda}}
- i\pi\log \frac{1 + \sqrt{1 - \lambda}} {1 - \sqrt{1 - \lambda}}
~~{\rm for}~~ \lambda < 1 
\label{eqn:spence}
\eea
We find that only the KK excitations of the top quark has a relevant 
contribution to the partial decay width. The UED contributions include the 
sum over the KK towers of the respective particle. As the more massive modes 
in the loop will hardly make significant contributions, we ensure that the 
sum is terminated as the higher modes decouple. We include the corrections 
to all the decay modes affected by UED contributions in the decay package 
HDECAY \cite{hdecay} to evaluate the relative sensitivities to the branching 
ratios to the different decay channels of the Higgs boson.

\begin{figure}[htb]
\begin{center}
\includegraphics[width=3.2in,height=3.2in]{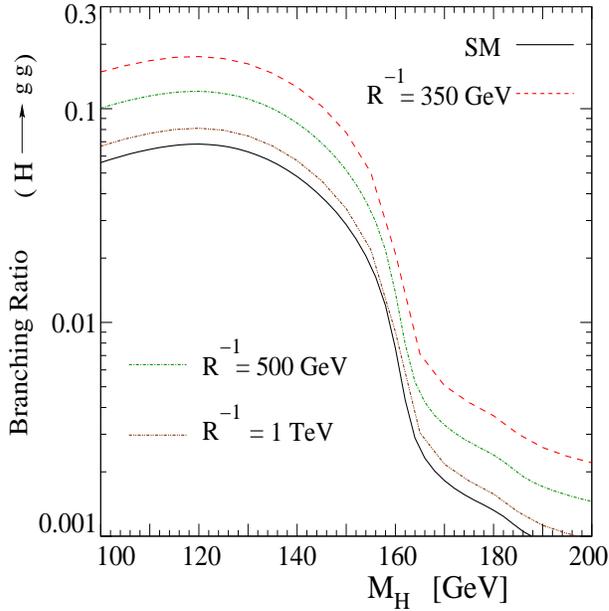}
\caption{ \it Illustrating the effects of UED contributions on the branching
ratio of $h \to gg$ as a function of Higgs mass ($M_H$) for different values
of the inverse of radius of compactification $R^{-1}$.}
\label{bratio}
\end{center}
\end{figure}

To illustrate the effect of the KK-modes on the branching ratio of $H\to gg$,
we plot the UED enhanced result along with the SM result as a function of Higgs
mass in Figure~\ref{bratio} for different values of $R^{-1}$. One can clearly 
see that large effects from UED result for lower values of $R^{-1}$. However,
very low values of $R^{-1}$ are constrained by precision electroweak data. 

\section{Results and Discussions} 
We have considered the case of a 500~GeV linear $e^+e^-$
collider and calculated the production of a Higgs in association with
a $Z$ boson \cite{higgstrahlung1,higgstrahlung2,higgstrahlung3} through a 
process of the form
\bea 
e^+e^- & \to & Z + H \nonumber \\ 
& \hookrightarrow & \ell^+ \ell^- + H \nonumber 
\eea 

\bigskip\noindent
In the preceding paragraph we discuss the final states relevant for our
study. 
\begin{enumerate}
\item $e^+ e^- \to \ell^+ \ell^- +$ two jets, which arises when the
Higgs boson decays to a pair of quarks ($u\bar{u},d\bar{d},c\bar{c},s\bar{s}$ 
and $b\bar{b}$) or gluons\footnote{We
exclude $H \to \tau^\pm$ decays because these produce narrow jets which can
be identified as $\tau^\pm$ with 80-90\% efficiency.}, which then
undergo fragmentation to form a pair of hadronic jets. Clearly, for a
Higgs boson in the SM,  final state will receive contributions 
mainly from the decays $H \to b\bar b$ and $H \to c \bar c$, with a 
minuscule contribution due to $H \to gg$. However, due to the increase
in the partial width for $H \to gg$ due to the extra contribution
coming from the additional KK excitations of the top quark in the loops, 
there will be an enhancement in the overall branching ratio to jets.

\item $e^+ e^- \to \ell^+ \ell^- + b \bar b$, which simply means that
the final state in the above contains two tagged $b$-jets. The decay
width for $H \to b \bar b$ is roughly the same in SM as well as UED, 
although the change of the two-gluon decay mode will have a small effect on 
the branching ratio for the $b \bar b$ mode. 
This would be observable in the rates for $b \bar{b}$ final state only
if the change in the two-gluon mode is quite large, considering the fact
that the corresponding branching ratios differ by more than an order of
magnitude in the intermediate mass range of the Higgs boson. However, this 
small change in the branching ratio of $H \to b \bar b$ can still play 
a pivotal role in identifying effects of UED, as we show in our analysis. 
\end{enumerate}

\bigskip\noindent
In our subsequent analysis, we have imposed a few kinematic acceptance 
cuts on the final state particles, viz.,
\begin{enumerate}
\item The final-state leptons should have transverse momentum
$p_T^{(\ell)} > 10$~GeV.
\item The final-state leptons should have pseudo-rapidity
$\eta^{(\ell)} < 3.0$.
\item The final-state jets should be clearly separated from each oher.
This can be implimented in a parton level analysis like ours by imposing 
a cut: $\Delta R_{JJ} (\equiv \sqrt{\Delta \eta_{JJ}^2 +
\Delta \phi_{JJ}^2}) > 0.4$, which is the usual criterion adopted at,
for example,
the LEP and Tevatron colliders.
\item The final-state jets should have transverse momentum $p_T^{(J)} >
10$~GeV.
\item The final-state jets should have pseudo-rapidity $\eta^{(J)} <
2.5$.
\end{enumerate}
The $b$-tagging efficiency \cite{desch} has been taken to be 50\%, which 
is probably a conservative estimate. 

\begin{figure}[htb]
\begin{center}
\includegraphics[height=3.3in,width=3.2in]{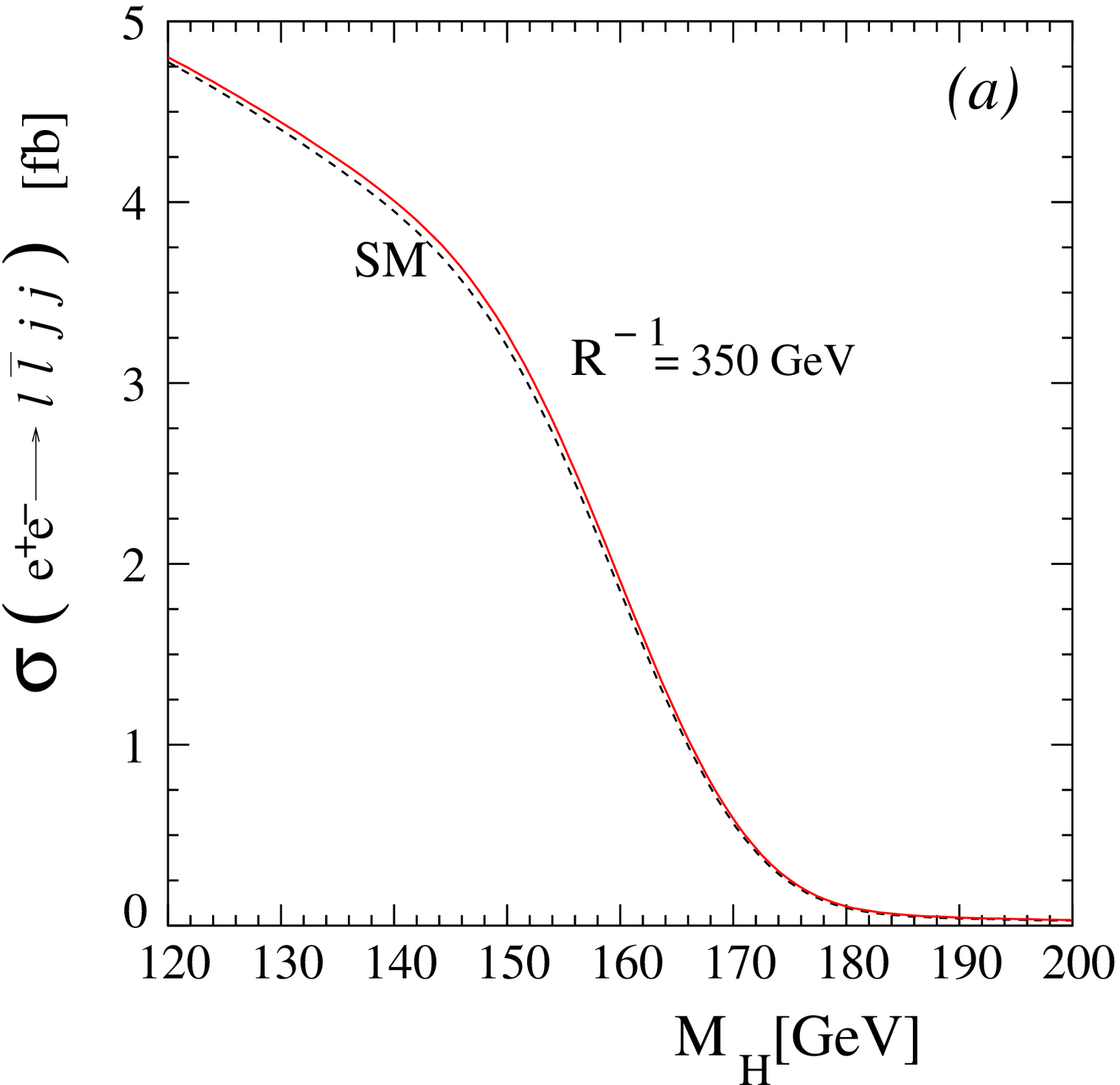}
\includegraphics[height=3.2in,width=3.25in]{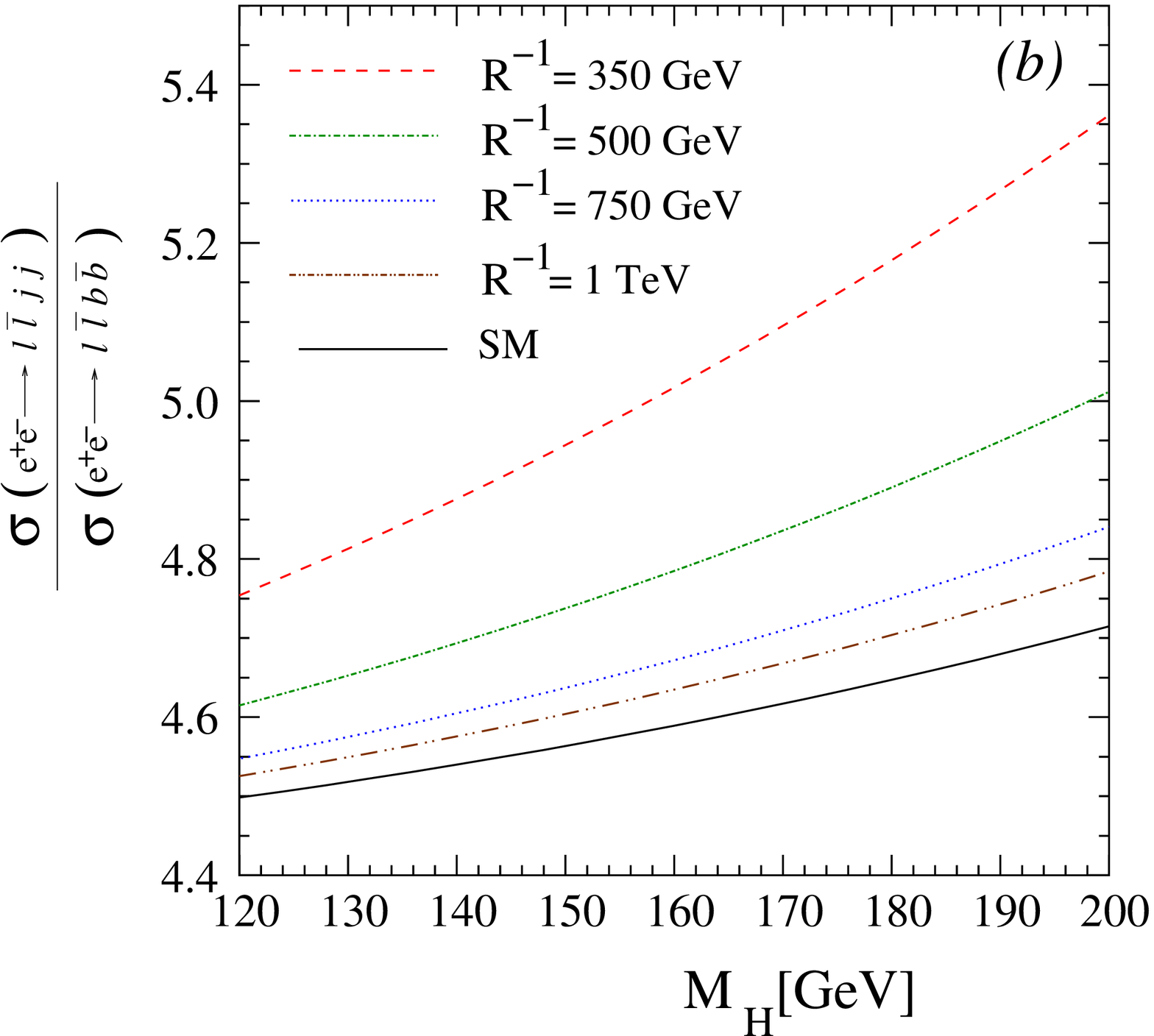}
\caption{\sl\small The curves are generated for $\sqrt{s}=500$ GeV
linear $e^+e^-$ collider.(a) Shows the cross-section for the process
($e^+ e^- \to \ell^+\ell^- ~+~ two~jets$) where the solid (red) line
includes the UED contribution while the dashed (black) line
corresponds to the SM prediction. (b) Shows the ratio of cross-section
between two final states viz. $\frac{\s(e^+e^- \to \ell^+\ell^- ~+~
two ~jets)}{\s(e^+e^- \to \ell^+\ell^- + b \bar b)}$ for different
values of the compactification radius. The solid (black) line
corresponds to the SM prediction.}
\label{fig1}
\end{center}
\end{figure}

\bigskip\noindent
In Figure~\ref{fig1}(a), we illustrate our result for the process discussed
above, namely, 
 $$ e^+e^- \to \ell^+\ell^- ~+~ two ~jets $$
at a $\sqrt{s}=500$ GeV, $e^+e^-$ collider.
The solid (red) line denotes the UED-included cross-section where we
have chosen the value of $R^{-1} = 350$ GeV which gives a greater
enhancement compared to values of $R^{-1}$ greater than the above, while
the dashed (black) line denotes the SM contribution only.  It should
be noted that the graph shows the excess cross-section after removing
the non-Higgs part of the Standard Model contributions (such as
$e^+e^- \to ZZ, ZZ^*$ etc.). These lead to a large SM four-fermion
background, which is however, easily reducible by selecting only
events corresponding to peaks in the $\ell^+\ell^-$ around $Z$-mass and
dijet invariant masses, but rejecting those events with 2-jet
invariant mass corresponds to $Z$-mass. The continuum background
($\gamma^\ast \gamma^\ast$, $Z^\ast Z^\ast$) too can be easily
neglected as it lies below $10^{-3}~fb$ (in the bins of $b \bar b$
invariant mass) and would hardly affect the rates for the signal in
consideration. The cross-section shown in Figure~\ref{fig1}(a) makes
it clear that it is very hard to see for differences by just looking
at the rates. The cross-sections for $\ell^+\ell^- + {\rm two~jets}$
final state are almost identical in the two cases.  As the
cross-sections look very similar, it would require very precise
measurements to form a distinction between the two cases.

\bigskip\noindent Thus, it seems that the two jet decay mode of the
Higgs alone will not be able to show observable difference due to the
enhancement in the decay channel of the two-gluon mode due to UED
contributions.  However if we consider the ratio of the two processes
mentioned above, viz.  $\frac{\s(e^+e^- \to \ell^+\ell^- ~+~ two
  ~jets)}{\s(e^+e^- \to \ell^+\ell^- + b \bar b)}$ we can see that the
difference between the two cases becomes more prominent. 
The number of 2$\ell +$ 2 $jets$ events do not differ in SM and 
UED to a great extent as revealed from Figure~\ref{fig1}(a). However, 
efficacy of the above mentioned ratio is also evident from the following 
figure (Fig1~\ref{fig1}(b)). Enhancement of the gluonic width of the Higgs 
boson in UED case has two fold effect on the ratio. It not only increases the 
$gg$ branching ratio (which affects the numerator of the ratio), but also 
suppresses the branching ratio in $b\bar b$-channel (thus suppressing the 
denominator and hence amplifies the ratio).
In Figure~\ref{fig1}(b) we plot this ratio for different values of the
compactification scale $R^{-1}$. We find that the ratio differs from
that of the SM throughout the mass range of $120~{\rm GeV} \leq m_H
\leq 200~{\rm GeV}$ with the lines converging towards the SM value as
$R^{-1}$ is increased. In fact this also highlights the decoupling
nature of the higher levels of the KK tower and justifies our
termination of the sum of KK towers in the loop, to values where the
contributions become negligibly small. The ratios tend to diverge more
as the Higgs mass increases. This is because the branching ratios of
$H\to gg$ and $H\to b\bar{b}$ become comparable and the enhancement in
the $H\to gg$ mode starts playing a more significant role in the 2-jet
final state. However, there is a caveat.  For comparatively higher
Higgs masses, cross-sections for both the above processes are
small. Hence, we need higher luminosity to differentiate between the
SM and the UED cases in a statistically significant way.  The
robustness of this method is, nevertheless, highlighted in the fact
that although the $H\to gg$ branching ratio is more than an order
smaller than that of $H\to b\bar{b}$ in the intermediate mass range
for the Higgs boson, we are still able to identify the difference due
to the UED contribution which would have been otherwise very difficult
to see, by just looking at the cross-sections.

\bigskip\noindent
It is necessary to mention here that the differences in the ratios as
evident from the graphs is due to the difference in the corresponding
branching ratios. The ratios are not susceptible to uncertainties like
different efficiency factors associated with particle identifications
as they would cancel out. The efficiency factors will however give a more
realistic estimate of the events that will be observed at the
experiments and which gives us an estimate of the uncertainties in the
statistics. 
 

\bigskip\noindent A similar exercise can be carried out for a 1 TeV
machine too, but the rates for the above processes would rapidly fall
due to the s-channel suppression. Infact, we find that the
cross-section for the $\ell^+\ell^- +~{\rm two~jets}$ process falls by
an order of magnitude compared to that of $\sqrt{s}=500$ GeV collider.
A better option at the higher $\sqrt{s}$ machine would be the
production of Higgs through the $WW-$fusion channel with the final
states being ``${\rm two~jets} + E\slash{}_T$" and ``$b \bar b +
E\slash{}_T$". The ratio between the final states will again show
differences, but it would be challenging to isolate the signal from
the large continuum background for the above process without losing
out much on the signal. Since the major background will be due to the
$e^+ e^- \to \g ^\ast Z,\g ^\ast Z^*$, etc. a large proportion of that
can be suppressed by rejecting events for $M_{inv}^{missing} < 100$
GeV.\footnote{$M_{inv}^{missing}$ is defined to be the mass of the
  system recoiling against the jet-pairs} Due to the t-channel nature
of the background processes, the momentum vector reconstructed from
the two jets, would be mostly in the high rapidity region, in contrast
to the same momentum vector coming from the Higgs boson produced by
the $WW-$fusion process. The distribution of the latter, would be
expected to peak in the central rapidity region. This too can provide
for a substantial suppression of the background by putting an
appropriate kinematic cut. However, we should accept the possibility
that it might just be kinematically allowed, at a $\sqrt{s}=1$ TeV
machine, to pair produce the n=1 KK excitations for direct
observation. In which case our results for the $\sqrt{s}=500$ GeV
machine becomes all the more interesting and worthy of attention.

\bigskip\noindent To summarise, we have shown that the effects of KK
excitations in the UED scenario can be evident at future lepton
colliders, much before the direct observation of these particles
(which are constrained to be produced in pairs), through the decay
mode of Higgs, driven by loops.  We find that, although these
enhancements might not be distinguishable by just looking at a single
cross-section viz. $\ell^+\ell^- +~{\rm two~jets}$, this difference
can be highlighted by looking at the ratio of cross-sections for two
different final states. Another advantage of taking ratios is that
many systematic errors and efficiency factors cancel out, making it a
more useful option. Infact, considering ratios to determine parameters
have been a usual practice where large uncertainties and errors are
involved. We also find that the differences are more pronounced for
higher Higgs masses, but are constrained by the low cross-sections for
the specific process in consideration, in that mass range. However,
the high luminosity achievable at the future lepton colliders is expected 
to probe this mass range too.

\bigskip\noindent
In fact the above technique can be used to identify other new physics
scenarios predicting massive particles, 
which play a similar role in modifying the partial width of the
Higgs to massless gauge bosons. It can also be used to distinguish scalars
of other theories which behave similar to the Higgs boson, like the Radion which has 
similar couplings like the Higgs boson\cite{skrai}. A major difference is the
enhanced coupling of Radion to gluons through the trace anomaly. 

\vskip 5pt
{\bf Acknowledgments:}
The authors would like to thank Anirban Kundu for useful discussions.
The authors would also like to thank the  organisers of "Study Group on 
Extra dimensions at LHC", held in June, 2005 at Harish-Chandra Research
Institute, where the early part of this work was initiated.


\begin{thebibliography}{99}

\bibitem{LED1}
  N.~Arkani-Hamed, S.~Dimopoulos and G.~R.~Dvali,
  Phys.\ Lett.\ B {\bf 429}, 263 (1998);
\bibitem{LED2}
  I.~Antoniadis, N.~Arkani-Hamed, S.~Dimopoulos and G.~R.~Dvali,
  Phys.\ Lett.\ B {\bf 436}, 257 (1998).

\bibitem{RS1}
  L.~Randall and R.~Sundrum,
  Phys.\ Rev.\ Lett.\  {\bf 83}, 3370 (1999);
\bibitem{RS2}
  L.~Randall and R.~Sundrum,
  Phys.\ Rev.\ Lett.\  {\bf 83}, 4690 (1999).

\bibitem{antoniadis1}
  I.~Antoniadis,
  Phys.\ Lett.\ B {\bf 246}, 377 (1990).

\bibitem{appel}
  T.~Appelquist, H.~C.~Cheng and B.~A.~Dobrescu,
  Phys.\ Rev.\ D {\bf 64}, 035002 (2001).
\bibitem{appel-yee}
T.~Appelquist and H.~U.~Yee,
  Phys.\ Rev.\ D {\bf 67}, 055002 (2003).

\bibitem{phenolow1}
  K.~Agashe, N.~G.~Deshpande and G.~H.~Wu,
  Phys.\ Lett.\ B {\bf 514}, 309 (2001);
\bibitem{phenolow2}
  D.~Chakraverty, K.~Huitu and A.~Kundu,
  Phys.\ Lett.\ B {\bf 558}, 173 (2003);
\bibitem{phenolow3}
  A.~J.~Buras, M.~Spranger and A.~Weiler,
  Nucl.\ Phys.\ B {\bf 660}, 225 (2003);
\bibitem{phenolow4}
  A.~J.~Buras, A.~Poschenrieder, M.~Spranger and A.~Weiler,
  Nucl.\ Phys.\ B {\bf 678}, 455 (2004);
\bibitem{phenolow5}
  J.~F.~Oliver, J.~Papavassiliou and A.~Santamaria,
  Phys.\ Rev.\ D {\bf 67}, 056002 (2003);
\bibitem{phenolow6}
  T.~Appelquist and B.~A.~Dobrescu,
  Phys.\ Lett.\ B {\bf 516}, 85 (2001);
\bibitem{phenolow7}
  K.~Agashe, N.~G.~Deshpande and G.~H.~Wu,
  Phys.\ Lett.\ B {\bf 511}, 85 (2001).

\bibitem{phenohigh1}
  T.~G.~Rizzo and J.~D.~Wells,
  Phys.\ Rev.\ D {\bf 61}, 016007 (2000);
\bibitem{phenohigh2}
  A.~Strumia,
  Phys.\ Lett.\ B {\bf 466}, 107 (1999);
\bibitem{phenohigh3}
  C.~D.~Carone,
  Phys.\ Rev.\ D {\bf 61}, 015008 (2000);
\bibitem{phenohigh4}
  C.~Macesanu, C.~D.~McMullen and S.~Nandi,
  Phys.\ Rev.\ D {\bf 66}, 015009 (2002);
\bibitem{phenohigh5}
  C.~Macesanu, C.~D.~McMullen and S.~Nandi,
  Phys.\ Lett.\ B {\bf 546}, 253 (2002);
\bibitem{phenohigh6}
  H.~C.~Cheng,
  Int.\ J.\ Mod.\ Phys.\ A {\bf 18}, 2779 (2003);
\bibitem{phenohigh7}
  A.~Muck, A.~Pilaftsis and R.~Ruckl,
  Nucl.\ Phys.\ B {\bf 687}, 55 (2004);
\bibitem{phenohigh8}
  G.~Bhattacharyya, P.~Dey, A.~Kundu and A.~Raychaudhuri,
  arXiv:hep-ph/0502031;
\bibitem{phenohigh9}
  M.~Battaglia, A.~Datta, A.~De Roeck, K.~Kong and K.~T.~Matchev,
  JHEP {\bf 0507}, 033 (2005);
\bibitem{phenohigh10}
  H.~C.~Cheng, K.~T.~Matchev and M.~Schmaltz,
  Phys.\ Rev.\ D {\bf 66}, 056006 (2002);
\bibitem{phenohigh11}
  B.~Bhattacherjee and A.~Kundu,
  arXiv:hep-ph/0508170;
\bibitem{phenohigh12}
  J.~M.~Smillie and B.~R.~Webber,
  arXiv:hep-ph/0507170;
\bibitem{phenohigh13}
  M.~Kakizaki, S.~Matsumoto, Y.~Sato and M.~Senami,
  Phys.\ Rev.\ D {\bf 71}, 123522 (2005).

\bibitem{petriello}
  F.~J.~Petriello,
  JHEP {\bf 0205}, 003 (2002).

\bibitem{pdfscale1}
  A.~Djouadi and S.~Ferrag,
  Phys.\ Lett.\ B {\bf 586}, 345 (2004);
\bibitem{pdfscale2}
  V.~Ravindran, J.~Smith and W.~L.~van Neerven,
  Nucl.\ Phys.\ B {\bf 665}, 325 (2003).

\bibitem{LKP1}
  G.~Servant and T.~M.~P.~Tait,
  Nucl.\ Phys.\ B {\bf 650}, 391 (2003);
\bibitem{LKP2}
  D.~Majumdar,
  Phys.\ Rev.\ D {\bf 67}, 095010 (2003).

\bibitem{branebulk1}
  H.~Georgi, A.~K.~Grant and G.~Hailu,
  Phys.\ Lett.\ B {\bf 506}, 207 (2001);
\bibitem{branebulk2}
  G.~von Gersdorff, N.~Irges and M.~Quiros,
  Nucl.\ Phys.\ B {\bf 635}, 127 (2002);
\bibitem{branebulk3}
  H.~C.~Cheng, K.~T.~Matchev and M.~Schmaltz,
  Phys.\ Rev.\ D {\bf 66}, 036005 (2002);
\bibitem{branebulk4}
  M.~Puchwein and Z.~Kunszt,
  Annals Phys.\  {\bf 311}, 288 (2004).

\bibitem{hdecay}
  A.~Djouadi, J.~Kalinowski and M.~Spira,
  Comput.\ Phys.\ Commun.\  {\bf 108}, 56 (1998)
  [arXiv:hep-ph/9704448].


\bibitem{higgstrahlung1}
  J.~R.~Ellis, M.~K.~Gaillard and D.~V.~Nanopoulos,
  Nucl.\ Phys.\ B {\bf 106}, 292 (1976);
\bibitem{higgstrahlung2}
  B.~W.~Lee, C.~Quigg and H.~B.~Thacker,
  Phys.\ Rev.\ Lett.\  {\bf 38}, 883 (1977);
\bibitem{higgstrahlung3}
  B.~L.~Ioffe and V.~A.~Khoze,
  Sov.\ J.\ Part.\ Nucl.\  {\bf 9}, 50 (1978)
  [Fiz.\ Elem.\ Chast.\ Atom.\ Yadra {\bf 9}, 118 (1978)].

\bibitem{desch}
See for example, K.~Desch, "Particle searches at a Linear Collider,
ICHEP2000, Osaka, 2000.

\bibitem{skrai}
  P.~K.~Das, S.~K.~Rai and S.~Raychaudhuri,
  Phys.\ Lett.\ B {\bf 618}, 221 (2005).
 [arXiv:hep-ph/0410244].
\end{thebibliography}
\end{document}